\documentclass{emulateapj}

\usepackage{graphicx,amsmath}
\usepackage{color}
\usepackage{times}

\def\mbh{M_{\bullet}}
\def\mbhp{M_{\bullet,\rm p}}
\def\mbhs{M_{\bullet,\rm s}}
\def\teff{T_{\rm eff}}
\def\rg{R_{\rm g}}
\def\sigB{\sigma_{\rm B}}
\def\hp{h}
\def\abbh{a_{\rm BBH}}
\def\msun{M_{\odot}}
\def\kms{\rm km~s^{-1}}
\def\FJ{F_{J}}
\def\feii{Fe\,{\footnotesize{II}}}

\def\be{\begin{equation}}
\def\ee{\end{equation}}
\def\bea{\begin{eqnarray}}
\def\eea{\end{eqnarray}}

\newcommand{\ltsim}{\mathrel{\protect\raisebox{-0.5ex}{$\:\stackrel{\textstyle
<} {\sim}\:$}}}
\newcommand{\gtsim}{\mathrel{\protect\raisebox{-0.5ex}{$\:\stackrel{\textstyle
>} {\sim}\:$}}}

\shorttitle{Supermassive Binary Black Hole in Mrk\,231}
\shortauthors{Yan et al.}

\begin{document}

\title{
A probable Milli-Parsec Supermassive Binary Black Hole in the Nearest
Quasar Mrk\,231
}
\author{Chang-Shuo Yan$^1$, Youjun Lu$^{1,\dagger}$, Xinyu Dai$^2$, \&
Qingjuan Yu$^3$ 
%\\
}
\affil{
%\normalsize{
$^1$National Astronomical Observatories, Chinese Academy of Sciences,
Beijing, 100012, China; $^\dagger$\,luyj@nao.cas.cn \\
%} \\
%
%\normalsize{$^2$
$^2$Homer L. Dodge Department of Physics and Astronomy, The University of
Oklahoma, Norman OK, 73019, USA \\
%} \\
%
%\normalsize{
$^3$Kavli Institute for Astronomy and Astrophysics, Peking University,
Beijing, 100871, China %} } } }
}

\begin{abstract}
Supermassive binary black holes (BBHs) are unavoidable products of
galaxy mergers and are expected to exist in the cores of many quasars.
Great effort has been made during the past several decades to search
for BBHs among quasars; however, observational evidence for BBHs
remains elusive and ambiguous, which is difficult to reconcile with
theoretical expectations. In this paper, we show that the distinct
optical-to-UV spectrum of Mrk 231 can be well interpreted as emission
from accretion flows onto a BBH, with a semimajor axis of
$\sim590$\,AU and an orbital period of $\sim1.2$ year.  The flat
optical and UV continua are mainly emitted from a circumbinary disk
and a mini-disk around the secondary black hole (BH), respectively;
and the observed sharp drop off and flux deficit at
$\lambda\sim4000-2500$\AA\ is due to a gap (or hole) opened by the
secondary BH migrating within the circumbinary disk.  If confirmed by
future observations, this BBH will provide a unique laboratory to
study the interplay between BBHs and accretion flows onto them.  Our
result also demonstrates a new method to find sub-parsec scale BBHs by
searching for deficits in the optical-to-UV continuum among the
spectra of quasars.
\end{abstract}


\section{Introduction}\label{sec:intro}
Supermassive binary black holes (BBHs) are natural products of the
hierarchical mergers of galaxies in the $\Lambda$CDM cosmology and are
expected to be abundant \citep[e.g.,][]{BBR80, Yu02, MM05}, since many
galaxies (if not all) are found to host a supermassive black hole
(SMBH) at their centers \citep[e.g.,][]{Magorrian98, KormendyHo13}.
Evidence has been accumulated for SMBH pairs in active galactic nuclei
(AGNs) and quasars with perturbed galaxy morphologies or other merger
features \citep[e.g.,][]{Komossa03, Liu10, Comerford11, Fu12}.  These
SMBH pairs will unavoidably evolve to closely bound BBHs with
separations less than $1$\,pc. However, the evidence for BBHs at the
sub-pc scale is still elusive \citep[e.g.,][]{Popovic12}, which raises
a challenge to our understanding of the BBH merger process and the
formation and evolution of SMBHs and galaxies.

A number of BBH candidates in quasars have been proposed according to
various spectral or other features, such as the double-peaked,
asymmetric, or offset broad line emission \citep[e.g.,][]{BL09,
Tsalmantza11, Eracleous12, Ju13, Liu14}, the periodical variations
\citep[e.g.,][]{Valtonen08, Graham14}, and etc.; however, most of
those candidates are still difficult to be confirmed. Thus, it is of
great importance to find other ways to select and identify BBHs in
quasars.  Recently, \citet[][]{GM12} proposed that the continuum
emission from a BBH-disk accretion system, with unique observable
signatures between $2000$\AA\ and 2$\mu$m because of a gap or a hole
in the inner part, can be used to diagnose BBHs (see
\citealt{Sesana12, Roedig14, Yan14}, but \citealt{Farris14}). This
method may be efficient in identifying BBHs since many AGNs and
quasars have multi-wavelength observations and broad band spectra.
Those previous investigations only focus on theoretical predictions,
and the present paper is the first attempt to apply this method to fit real
observations.

In this paper, we report a BBH candidate in the core of Mrk~231, the
nearest quasar with a redshift $z=0.0422$, according to its unique
optical-UV spectrum.  In Section~\ref{sec:mrk231}, we summarize the
multi-wavelength spectrum of Mrk~231 and its distinctive spectral
features comparing with normal quasars. The spectrum of Mrk\,231 at
the optical band is similar to the quasar composite spectrum; however,
it drops dramatically at the wavelengths around 3000\AA\ and becomes
flat again at $\la 2500$\AA. This anomalous continuum is hard to be
explained by normal extinction/absorption \citep{Veilleux13}. We
propose that the unique optical-to-UV spectrum of Mrk\,231 can be
explained by emission from a BBH accretion system, with which the drop
of the continuum at $\la$ 4000\AA\ is due to a gap or a hole opened by
the secondary component of the BBH. In Section~\ref{sec:bbhmodel}, we
introduce a simple (triple-)disk model for the accretion onto a BBH
system.  Using this model, we fit the optical-to-UV continuum of
Mrk~231 (Section~\ref{sec:ouv}) and constrain the orbital
configuration of the BBH system and the associated physical parameters
of the accretion process in Section~\ref{sec:fitresults}.  Discussions
and conclusions are given in Sections~\ref{sec:discussions}  and
\ref{sec:condis}.

\section{Multi-band Observations of Mrk~231}\label{sec:mrk231}

Mrk\,231 is an ultraluminous infrared galaxy with a bright quasar-like
nucleus. It is probably at the final stage of a merger of two galaxies
as suggested by its disturbed morphology and the associated tidal
features \citep{Lipari94, Armus94}. The broadband spectrum of the
Mrk\,231 nucleus exhibits some extreme and surprising properties as
follows.

First, the flux spectrum ($F_{\lambda}$) drops dramatically by a
factor of $\sim 10$ at  the near UV band (from wavelength $\lambda
\sim 4000$\,\AA\  to $2500$\,\AA), while it is flat at $\lambda \sim
1000$\,\AA - $2500$\,\AA\ and at $\lambda \sim
4000$\,\AA-$10000$\,\AA.  If this sharp drop off is due to extinction,
it requires a large dust reddening of $\rm A_{v} \sim 7$\,mag  at
$\lambda \sim 2500-4000$\,\AA\ and a small dust reddening $\sim
0.5$\,mag at $< 2500$\,\AA\ \citep{Veilleux13}.  Thus, an unusual
complex geometric structure for the extinction material surrounding
the disk must be designed \citep{Veilleux13, Leighly2014}. 

Second, Mrk\,231 is an extremely powerful Fe\,{\footnotesize{II}}
emitter as suggested by the spectrum on the blue side of H$_{\alpha}$
and on both sides of H$_{\beta}$. Such a high level of optical Fe line
emission suggests a significant amount of Fe line emission at the UV
band; however, it is not visible in the observed UV spectrum
\citep{Veilleux13, Leighly2014}.

Third, a number of broad low-ionization absorption line systems, such
as Na\,{\footnotesize I} D, Ca\,{\footnotesize II}, Mg\,{\footnotesize
II}, Mg\,{\footnotesize I}, and Fe\,{\footnotesize II}, have been
identified in the optical-to-UV spectrum, which suggests that Mrk~231
should be an Fe low-ionization broad absorption line quasar (FeLoBAL)
\citep{Veilleux13}. However, the expected corresponding absorption
features at the UV and FUV bands are not evident.  

Fourth, the hard X-ray emission of Mrk\,231 is extremely weak. The
intrinsic, absorption corrected, X-ray luminosity at $2$-$10$\,keV is
$L_{2-10{\rm keV}} \sim 3.8 \times 10^{42} \rm{erg~s^{-1}}$, and the
ratio of $L_{2-10{\rm keV}}$ to the bolometric luminosity (inferred
from the optical luminosity) is only $\sim 0.0003$ \citep{Saez12,
Teng14},  almost two orders of magnitude smaller than the typical
value  ($\sim 0.03$) of a quasar with a similar bolometric luminosity
\citep{Hopkins07}. 

Below we show that the above first feature, the anomalous UV
continuum, is a distinct prediction of a BBH--disk accretion system as
shown in Figure~\ref{fig:f1}. The last three features, commonly
interpreted under the context of complicated outflows with
absorptions, can be also accommodated under the framework of the
BBH--disk accretion, e.g., the wind features are consistent with the
Fe absorption features of a typical FeLoBAL, and Mrk\,231's intrinsic
X-ray weakness is also a natural consequence of a BBH--disk accretion
system with a small mass ratio.

\section{Optical-to-UV continuum from a binary black
hole---(triple-)disk accretion system}
\label{sec:bbhmodel}

\begin{figure}
\centering
\includegraphics[width=0.40\textwidth]{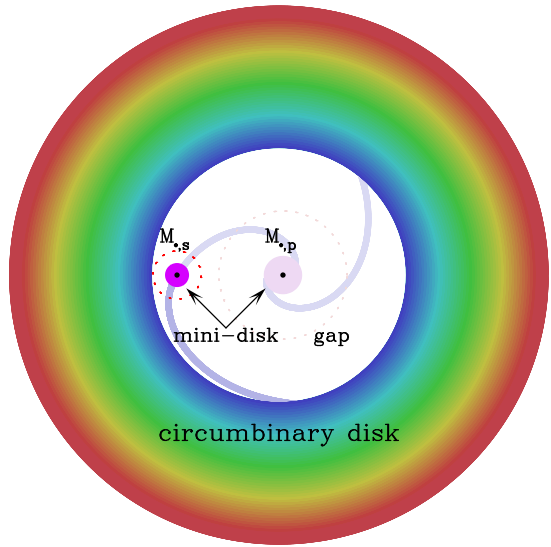}
\caption{
Schematic diagram for a BBH--disk accretion system.  The BBH is
assumed to be on circular orbits with a semimajor axis of $\abbh$, and
the masses of the primary and secondary components are
$M_{\bullet,{\rm p}}$ and $M_{\bullet,{\rm s}}$, respectively.  The
BBH is surrounded by a circumbinary disk, connecting with the
mini-disk around each component of the BBH by streams.  In between the
circumbinary disk and the inner mini-disks, a gap or hole is opened by
the secondary SMBH \citep{AL96,Dorazio13,Farris14b}.  The width of the
gap (or hole) is roughly determined by the Hill radius $R_{\rm H}$
[$\sim\abbh(M_{\bullet,{\rm s}}/3M_{\bullet,{\rm
p}})^{1/3}\simeq0.69q^{1/3}\abbh$], where $q$ is the mass ratio, and
the inner boundary of the circumbinary disk can be approximated as
$r_{\rm in,c}\sim\abbh/(1+q)+R_{\rm H}$.  The outer boundary of the
mini-disk surrounding the secondary SMBH ($r_{\rm out,s}$) is assumed
to be a fraction ($f_{\rm r,s}$) of the mean Roche radius, $R_{\rm
RL}(q)\simeq0.49\abbh q^{2/3}/[0.6q^{2/3}+\ln(1+q^{1/2})]$
\citep{Eggleton}, i.e., $r_{\rm out,s}=f_{\rm r,s}R_{\rm RL}(q)$,
considering that the mini-disk may not fill the whole Roche lobe (the
red dashed circle). For BBHs with mass ratios roughly in the range
from a few percent to $0.25$, the accretion onto the secondary SMBH
and consequently its emission dominates, compared with that from the
mini-disk around the primary BH \citep{Roedig12,Farris14b}.  
}
\label{fig:f1}
\end{figure}

Considering a BBH system resulting from a gas rich merger, the BBH is
probably surrounded by a circumbinary disk, and each of the two SMBHs
is associated with a mini-disk (see Fig.~\ref{fig:f1}). In between the
circumbinary disk and the inner mini-disks, a gap (or hole) is opened
by the secondary SMBH, which is probably the most distinct feature of
a BBH--disk accretion system, in analogy to a system in which a gap or
hole is opened by a planet migrating in the planetary disk around a
star \citep{Lin96,Quanz13}.  This type of geometric configurations for
the BBH--disk accretion systems has been revealed by many numerical
simulations and analysis \citep{AL96, Dorazio13, Hayasaki08, Escala05,
Cuadra09, Roedig14, Farris14b}.\footnote{The width of the gap (or
hole) is roughly determined by, but could be somewhat larger than, the
Hill radius $R_{\rm H}$. However, set a slightly large gap size, e.g.,
$1.2R_{\rm H}$, does not affect the results presented in this paper
significantly.} The continuum emission from disk accretion onto a BBH
may be much more complicated than that from disk accretion onto a
single SMBH, since the dynamical interaction between the BBH and the
accretion flow onto it changes the disk structure \citep{GM12,
Sesana12, Rafikov13, Roedig14, Yan14, Farris14}.  Nevertheless, we
adopt a simple model to approximate the continuum emission from a
BBH--disk accretion system as the combination of the emissions from an
outer circumbinary disk and an inner mini-disk around the secondary
SMBH, each approximated by multicolor black body radiation in the
standard thin disk model \citep{SS73,NT73}. The emission from the
mini-disk around the primary SMBH is insignificant for a BBH system
with a small mass ratio (roughly in the range of a few percent to
$0.25$) due to its low accretion rate as suggested by the state of the
art numerical simulations \citep{Roedig12,Farris14b}, thus its
emission can be neglected. Our analysis suggests that a large $q$
cannot lead to a good fit to the observations.

\subsection{ Emission from the circumbinary disk }
We choose a standard thin disk to approximate the temperature profile
of the circumbinary disk.  The structure and spectral energy
distribution (SED) of the circumbinary disk may be different from that
of a standard thin disk, especially at the region close to the inner
edge, since the torque raised by the central BBH may lead to gas
accumulation there. In this region, the circumbinary disk is somewhat
hotter than the corresponding region of a standard thin disk with the
same accretion rate and total SMBH mass ($M_{\bullet,{\rm p}} +
M_{\bullet,{\rm s}}$) \citep[e.g.,][]{Rafikov13}. We neglect this
slight difference for now and will discuss this in the Appendix.

\subsection{ Emission from the inner mini-disk associated with the
secondary SMBH } 
We assume that the emission from the inner mini-disks is dominated by
that from the mini-disk around the secondary SMBH, and the emission
from a mini-disk can also be approximated by that from a standard thin
disk with the same extent, accretion rate, and SMBH mass.  The
accretion onto each of the two SMBHs may be highly variable
\citep{Roedig12, Hayasaki08}, and the temperature structure of each of
the two mini-disks may be affected by the torque from the other SMBH
component, the outer circumbinary disk \citep{Farris14}, and the
infalling stream. We ignore those complications in the fitting and
will discuss the related effects in Appendix.

In the standard thin accretion disk model, the emission from an
annulus $r- dr/2\to r+dr/2$ of the disk is approximated by a black
body radiation with an effective temperature of
\bea
\teff(r) & = & \left[\frac{3GM_{\bullet} \dot{M}_{\rm acc}}{8\pi\sigB
r^3}\left(1-\sqrt{\frac{r_{\rm in}}{r}}\right)\right]^{1/4},
\label{eq:temp}
\eea
where $G$ is the gravitation constant, $\sigma_{\rm B}$ is the
Stefan-Boltzmann constant, $\dot{M}_{{\rm acc}}$ is the accretion rate
of the SMBH, and $r_{{\rm in}}$ is the radius of the disk's inner
edge. For the circumbinary disk, $r_{\rm in,c} = \abbh/(1+q)+R_{\rm
H}$; for the mini-disk around the secondary SMBH, $r_{{\rm in,c}} \sim
3.5 GM_{\bullet,{\rm s}} /c^2 = 3.5r_{\rm g,s}$, assuming a radiative
efficiency $\epsilon \simeq 0.1$ (correspondingly an SMBH spin of
$\simeq 0.67$; \citealt{YL08,Shankar13}). Here $c$ is the speed of
light.  We can then obtain the continuum emission for either the
circumbinary disk or the mini-disk around the secondary SMBH as
\be
F_{\lambda}=\int^{r_{{\rm out}}} _{r_{\rm in}}
\frac{2\pi r}{D_{\rm l}^2}
\frac{2 hc^2 \cos i
/\lambda^5}{\exp[{\hp c/\lambda k_{\rm B}\teff(r)}]-1}dr,
\label{eq:F1}
\ee
where $D_{\rm l}$ is the luminosity distance of Mrk\,231, $h$ is the Planck constant, $k_{\rm B}$ the Boltzmann constant,
$i$ the inclination angle, $r_{\rm out} = 10^5r_{\rm in,c}$ and
$f_{\rm r,s} R_{\rm RL}(q)$ for the circumbinary disk and the
mini-disk, respectively. Here we simply assume $\cos i = 0.8$, the
mean value for type 1 quasars. We find that a slight change of $\cos
i$ does not affect our results significantly.

\subsection{ Pseudo-continuum from Fe emission lines }
It has been shown that there are extremely strong \feii\, emission
lines in the optical spectrum of Mrk 231. Thousands of Fe emission
lines from the broad line emission region blending together can form a
pseudo-continuum. Therefore,  the Fe emission must be included when
fitting  the SED of Mrk~231. We use the template-fitting method
introduced by \citet{phil77}, probably a standard practice, to treat
the Fe emission in Mrk\,231, in which the Fe spectrum of the
narrow-line Seyfert 1 galaxy I\,Zw\,1 is used to construct an \feii\,
template. In the UV band, we adopt the Fe template of \citet{mv01}. In
the optical band, the Fe template is constructed according to the list
of the Fe lines for I~Zw~1 given in \citet{veron04}.  We combine the
UV and optical Fe templates to form a single template, convolve it by
a Gaussian function with a width of ${\rm FWHM_{\rm conv}}$, and scale
it by a factor of $A_{\rm s}$ to match the observations.  The total
flux of the Fe emission, $I_{\rm Fe}$, is equal to $A_{\rm s}$
multiplied by the total flux of the template; the FWHM of the Fe lines
can be expressed as ${\rm FWHM_{\rm Fe}}=\sqrt{{\rm FWHM_{I Zw
1}^2+FWHM_{\rm conv}^2}}$.  In the above algorithm, we assume that the
Fe emissions in the UV and optical bands have the same width and there
is no shift between the redshifts of the UV and the optical lines. We
also assume that the ratio of the UV Fe flux to the optical Fe flux is
fixed to be the same as that of I~Zw~1.  A detailed dynamical model
for the broadening and the shift of individual Fe emission lines is
beyond the scope of our paper.

We have also further checked the validity of adopting the Fe template
of I\,Zw\,1 by using a CLOUDY \citep{Ferland13} model, with the best
fit continuum of Mrk\,231 obtained below as the input intrinsic
continuum.  We find that the total flux of the UV Fe emission ($I_{\rm
Fe,UV}$) to the total flux of the optical Fe emission ($I_{\rm
Fe,opt}$) generated from the CLOUDY model ($\sim 3$) is similar to
that of I\,Zw\,1 if the ionization parameter is sufficiently small
($\sim 0.001$).  A small ionization parameter is compatible with the
best-fit parameters of the BBH--disk accretion system, since the broad
line emission region is bound to the primary SMBH, much more massive
than the secondary SMBH, while the ionizing photons are provided by
the mini-disk around the secondary SMBH. If $I_{\rm Fe,UV}/I_{\rm
Fe,opt}$ is substantially smaller than that of I\,Zw\,1, then the
FeLoBAL feature in Mrk\,231 would be much less significant than that
shown in Figures~\ref{fig:f3} and \ref{fig:f7}.

\section{The Optical-to-UV Spectrum of Mrk\,231}
\label{sec:ouv}
\begin{figure}
\centering
\includegraphics[angle=-90,width=0.45\textwidth]{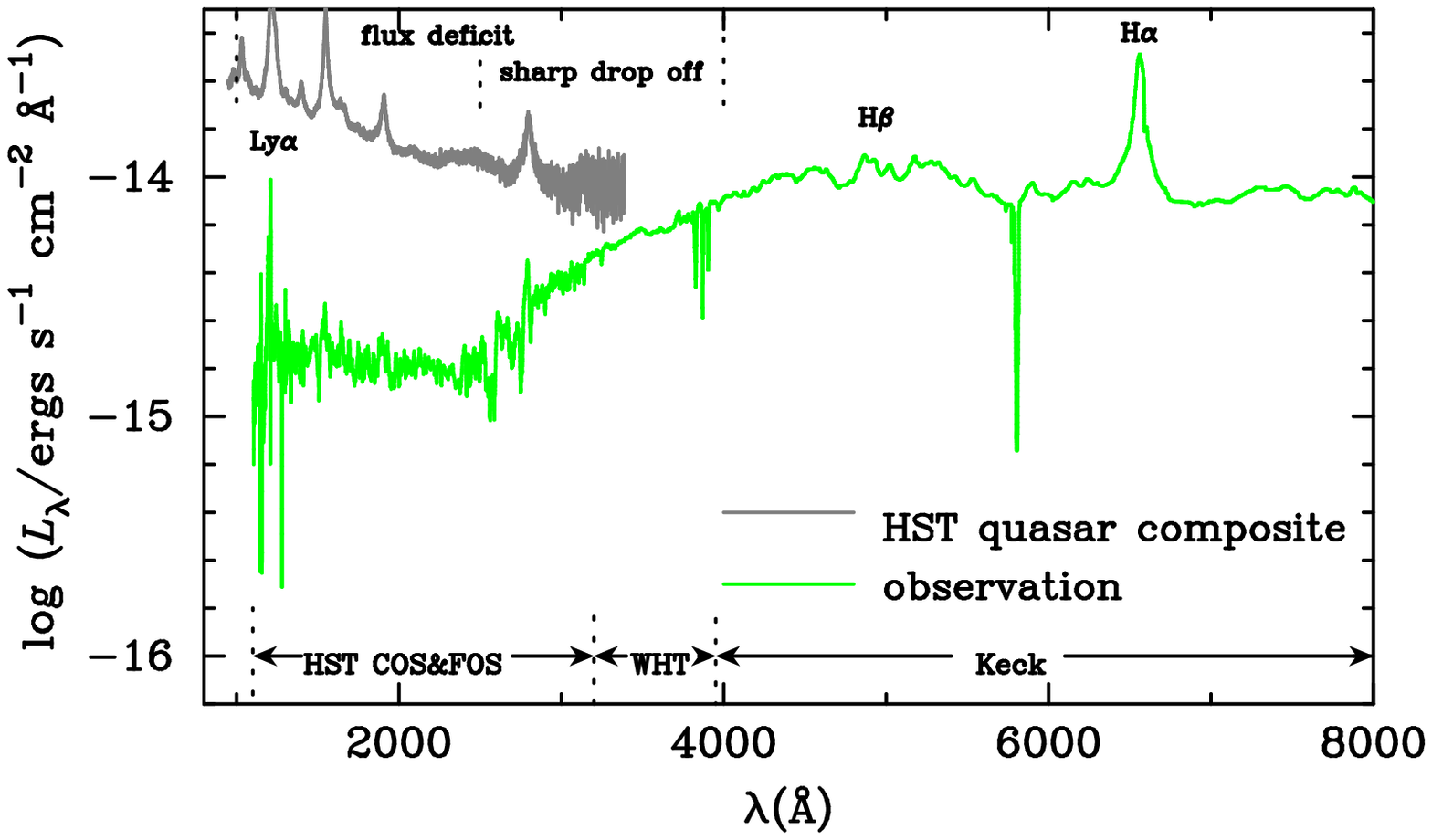}
\caption{
The optical-to-UV spectrum of Mrk231.  The green curve represents the
UV observations by the COS and FOS on board the HST (archive data;
\citealt{Veilleux13, Smith95}), the observations by the William
Herschel Telescope (WHT) and Keck telescope \citep{Leighly2014}. The
wavelength ranges for those different observations are marked at the
bottom of the figure. The FOS data is scaled up by a factor of $2$ to
match the COS data \citep[see][]{Veilleux13}. The grey curve
represents the HST composite spectrum of quasars \citep{Zhengw97}. The
continuum shows a sharp drop off at $\lambda \sim 4000-2500$\AA\ and
deficit of flux at $\lambda \sim 4000-1000$\AA. The locations of the
emission lines H$\alpha$, H$\beta$, and Ly$\alpha$ are also marked in
the figure. }
\label{fig:f2}
\end{figure}

\begin{figure}
\centering
\includegraphics[angle=-90,width=0.45\textwidth]{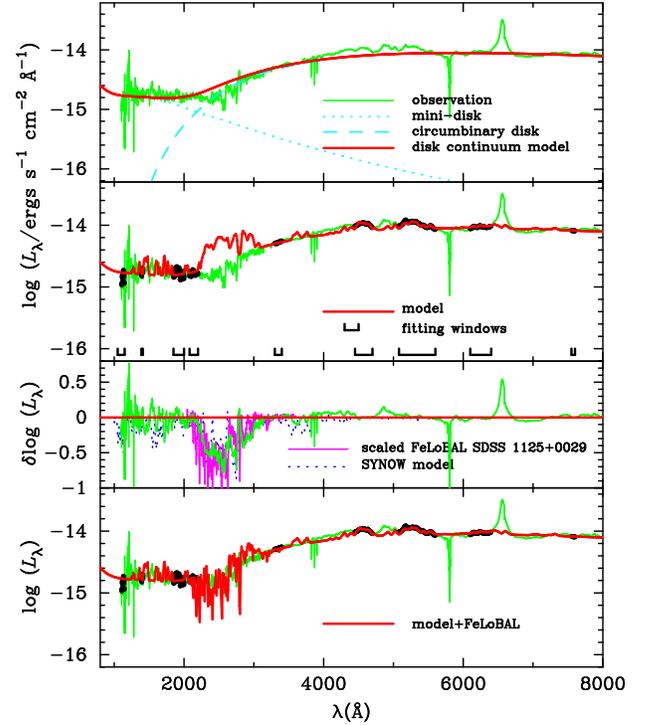}
\caption{
The optical-to-UV spectrum of Mrk231 and the model spectrum.  From top
to bottom, the first panel: the green curve represents the
observations as that in Figure~\ref{fig:f2}.  The red curve represents
the best-fit model for the continuum emission from the BBH--disk
accretion system, a combination of the continuum emissions from the
circumbinary disk (cyan dashed line) and the mini-disk around the
secondary SMBH (cyan dotted line).  The second panel: the green curve
represents the observational spectrum of Mrk\,231.  The red curve
represents the best-fit continuum spectrum, a combination of the
continuum emissions from the circumbinary disk, the mini-disk around
the secondary SMBH, and the pseudo continuum by Fe emissions. 
The third panel: the green curve represents the residuals of the
best fit continuum to the observations. The blue dotted curve
represents the Fe absorption obtained from a model using SYNOW code,
and the magenta solid curve represents the FeLoBAL absorption of SDSS
1125+0029 scaled by a factor of $1.6$.  The fourth panel: the red
curve represents the model spectrum by adding the pseudo-continuum due
to numerous Fe emission lines and including the contribution from the
FeLoBAL absorption. The black points represent the observational data
in those windows adopted in the continuum fitting as marked in this
panel.
}
\label{fig:f3}
\end{figure}

The observed broadband spectrum of Mrk~231 is first shifted to the
rest frame and then corrected for the Galactic extinction [$E({\rm
B-V})=0.02213$], and the results are shown in Figure~\ref{fig:f2}.  A
number of fitting windows are chosen in order to avoid those strong
emission or absorption lines, such as H$_\alpha$, H$_\beta$,
O\,{\footnotesize III}, Ly$\alpha$, and Na\,{\footnotesize I} D, which
is the commonly adopted way to perform the continuum or SED fitting.
The fitting windows adopted are $1140$-$1152$, $1260$-$1300$,
$1340$-$1400$, $1500$-$1650$, $1860$-$2000$, $3090$-$3145$,
$3740$-$3800$, $4400$-$4700$, $5080$-$5600$, $6060$-$6350$, and
$6750$-$6850$\AA. A more complicated model may be developed by fitting
each of the emission lines with one or more Gaussian components and
each of the absorption lines with Voigt profile or Gauss-Hermit
expansions; however, these lines will not affect the overall shape of
the continuum, the focus of this paper, and thus we choose to simplify
the presentation by not involving fittings to those emission and
absorption lines.  The wavelength range $3800$-$4400$\AA\, is also
excluded because of the contamination from the high-order Balmer
lines.  The wavelength range from $2000$ to $3000$\AA\, is excluded to
avoid the uncertainties in modeling strong Fe absorption lines as
indicated by the FeLoBAL signatures found in the optical band.

\section{MCMC fitting results}
\label{sec:fitresults}

We use the Markov Chain Monte Carlo (MCMC) method to obtain the best
fit to the observational data in the above fitting windows and
constrain the model parameters.  The parameters included in the
continuum model are the total mass ($M_{\bullet}$), mass ratio ($q$),
semimajor axis ($\abbh$) of the BBH, the Eddington ratios of the outer
circumbinary disk ($f_{\rm Edd,c}$) and the inner mini-disk ($f_{\rm
Edd,s}$), the ratio of the outer boundary of the inner mini-disk to
the  mean Roche radius ($f_{\rm r,s}$), the scale factor $A_{\rm s}$
and convolution width ${\rm FWHM_{\rm conv}}$ of the Fe emission
lines, and the extinction $E_{\rm B-V}$ due to the interstellar medium
in Mrk\,231.  Here the Eddington ratio is defined as the ratio of the
accretion rate of the circumbinary disk $\dot{M}_{\rm acc,c}$ (or the
secondary mini-disk $\dot{M}_{\rm acc,s}$) to the accretion rate
$\dot{M}_{\rm Edd}$ set by the Eddington limit (assuming
$\epsilon=0.1$), i.e., $f_{\rm Edd,c} = \dot{M}_{\rm acc,c}/
\dot{M}_{\rm Edd}(\mbh) = \epsilon \dot{M}_{\rm acc,c} c^2/L_{\rm
Edd}(\mbh)$, $f_{\rm Edd,s} = \dot{M}_{\rm acc,s}/\dot{M}_{\rm
Edd}(\mbhs) = \epsilon \dot{M}_{\rm acc,s} c^2/L_{\rm Edd}(\mbhs)$,
$L_{\rm Edd}(\mbh) =1.3\times 10^{46}{\rm erg~s^{-1}}
(\mbh/10^8\msun)$, and $L_{\rm Edd}(\mbhs) =1.3\times 10^{46}{\rm
erg~s^{-1}} (\mbhs/10^8\msun)$.  We first adopt those nine parameters
($M_{\bullet}$, $q$, $\abbh$, $f_{\rm Edd,c}$, $f_{\rm Edd,s}$,
$f_{\rm r,s}$, $A_{\rm s}$, ${\rm FWHM_{conv}}$, ${E_{\rm B-V}}$) to
fit the spectrum and obtain the best fit, and then fix  the two
parameters $A_{\rm s}$ and ${\rm FWHM_{conv}} (\sim 3000\kms)$ at
their best-fit values and obtain constraints on the other seven
parameters ($M_{\bullet}$, $q$, $\abbh$, $f_{\rm Edd,c}$, $f_{\rm
Edd,s}$, $f_{\rm r,s}$, ${E_{\rm B-V}}$). In the fitting, we adopt an
extinction curve for SMC according to \citep{Pei92};\footnote{We have
checked that choosing a different extinction curve (e.g., the one for
LMC) does not qualitatively affect our results.} the Eddington ratios
for the outer circumbinary disk and the inner mini-disk are assumed to
be in the range from $0.1$ to $1$, since the standard thin disk model
may be invalid if the Eddington ratio is substantially smaller than
$0.1$.

The top panel of Figure~\ref{fig:f3} shows the best-fit model to the
continuum emission from the BBH-disk accretion system.  The overall
shape of the observed continuum can be reproduced by the BBH-disk
accretion model.  The best fit parameters are $\mbh=1.5\times
10^8\msun$, $q=0.026$, $f_{\rm Edd,c}=0.5$, $f_{\rm Edd,s}=0.6$,
$\abbh= 590$\,AU$=2.9$mpc, and $f_{\rm r,s} = 0.11$,
respectively.\footnote{Note that the best fit value of $f_{\rm
Edd,s}=0.6$ means that the secondary SMBH accretes via a rate close to
the Eddington limit. The numerical simulations suggest that the accretion
rate onto the primary SMBH is smaller than that onto the secondary SMBH
for a BBH-disk accretion system with a mass ratio of a few to 25
percent \citep[e.g.,][]{Farris14b}. For the BBH system that best fits the
continuum, the accretion rate of the primary disk should be $\la 0.01$,
which is via the advection dominated accretion flow (ADAF) mode 
(\citealt{Esin97}) and radiatively extremely inefficient, and
thus its emission can be neglected. This validates the omission of
the primary disk emission in the fitting.  } According to the best-fit
model, the circumbinary disk dominates the emission of Mrk\,231 in the
optical band; the mini-disk dominates the FUV emission; the sharp drop
off at $4000-2500$\AA\, is mainly due to the cut off of the
circumbinary disk and the gap (or hole) opened by the secondary SMBH,
but not an extremely high extinction (see the top panel of
Figure~\ref{fig:f3}). It is also not a necessity to have different
extinctions in different bands to explain the sharp drop off. 

The second panel of Figure~\ref{fig:f3} also shows the best fit to the
continuum emission of Mrk\,231, which includes not only the continuum
emission from the circumbinary disk and the mini-disk around the
secondary SMBH, but also the pseudo continuum from numerous Fe
emission lines.  The Fe emission of Mrk\,231 in the optical band is
extremely high, and consequently the Fe emission in the UV band is
likely to be high, too. In the observed UV spectrum, however, there
appear no strong Fe emission lines \citep{Veilleux13, Leighly2014}.
Mrk\,231 is an FeLoBAL and should have very strong Fe absorptions in
the UV band as indicated by the optical absorption lines, such as
Na\,{\footnotesize{I}} D, but there appear no very strong Fe
absorption lines in the observed UV spectrum.  The reason is that the
Fe absorption dominates over the Fe emission in this band as detailed
below.

The third panel of Figure~\ref{fig:f3} shows the residuals of the
model spectrum at $2000$-$3200$\,\AA\, which reveals strong Fe
absorption lines. These Fe absorption features are quite similar to a
$1.6$-time scale up of the FeLoBAL absorption features of
SDSS~1125+0029 shown in Figure~4a of \citet[][the magenta
curve]{Hall02}, of which no observation at $\ltsim 2000$\AA\ is
available. As an illustration, we further use the SYNOW code
\citep{Branch02}, a fast parameterized synthetic-spectrum code, to
generate a spectrum for the absorption features in the NUV band for
Mrk\,231.  We assume that only the wind that covers the surface of the
disk facing toward distant observers contributes to the absorption
because of the optically thick disk, and this wind may launch at the
inner edge of the circumbinary disk and/or from the mini-disk.  For
simplicity, we consider three species, i.e., Fe {\footnotesize{I}}, Fe
{\footnotesize II}, and Mg {\footnotesize II}, and assume a
temperature of $10000$K, minimum velocity $v_{\rm min}=1000~\kms$, and
maximum velocity $v_{\rm max}=8000~\kms$. The maximum velocity is on
the same order of the escape velocity at $\abbh$.  The model
absorption spectrum is shown in the bottom panel of
Figure~\ref{fig:f3}. It appears that the main absorption features seen
in the residuals can be roughly modeled, though there are still some
discrepancies in details. In the bottom panel of Figure~\ref{fig:f3},
the model spectrum (red curve) is a combination of the best fit to the
continuum and a $1.6$-time scale up of the FeLoBAL absorption feature
of SDSS 1125+0029 (cyan line in Fig.~\ref{fig:f3}), which appears to
match the observations well.

\begin{figure}
\centering
\includegraphics[width=5.cm,angle=-90]{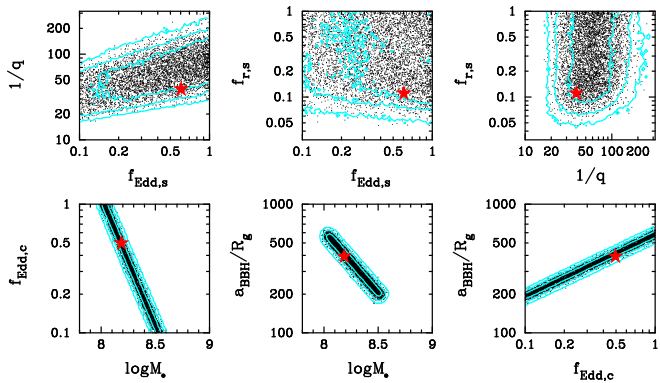}
\caption{
Two-dimensional probability contours for different parameters of the
BBH.  The red stars represent the best-fit parameter values.  Only
$50,000$ points (black) out of $50,000,000$ simulations are plotted.
The cyan curves represent 1-, 2-, and 3$\sigma$ confidence contours
from the inside to the outside, respectively. }
\label{fig:f4}
\end{figure}

\begin{figure}
\centering
\includegraphics[width=5.cm,angle=-90]{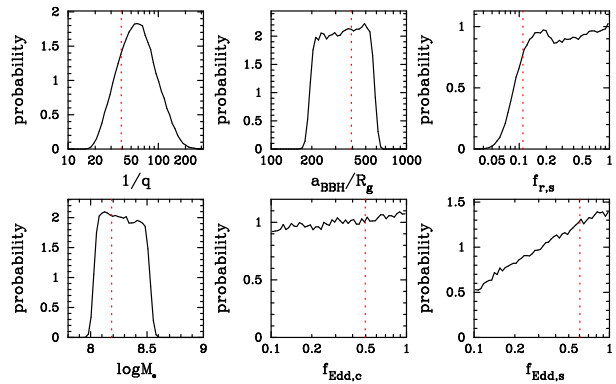}
\caption{
One-dimensional probability distributions of different parameters of
the BBH.  The red dotted vertical line in each panel represents the
value of the best fit adopted in Figs.~\ref{fig:f2} and \ref{fig:f3}
to produce the model spectrum.}
\label{fig:f5}
\end{figure}

Figures~\ref{fig:f4} and \ref{fig:f5} show the two-dimensional
probability contours and one-dimensional  marginalized probability
distributions for the model parameters obtained from the MCMC fitting,
respectively. The parameters $\mbh$, $f_{\rm Edd}$, and $\abbh$ mainly
determine the emission from the circumbinary disk and they are
strongly degenerate with each other. If one of them could be
determined by an independent method, the constraints on the other two
would be significantly improved. The emission from the mini-disk
around  the secondary SMBH is determined by $q$, $f_{\rm Edd,s}$, and
$f_{\rm r,s}$, for which $q$ and $f_{\rm Edd,s}$ are also degenerate
with each other. The constraints on these three parameters can be
substantially improved if observations at EUV (e.g., $\la 1000$\AA)
are available.  According to the marginalized one-dimensional
probability distribution for each parameter shown in
Figure~\ref{fig:f5} by the standard MCMC technique, we have the peak
values of the model parameters as $\log(\mbh/\msun) =
8.3_{-0.2}^{+0.2}$, $\log q = -1.8_{-0.2}^{+0.2}$, $\log f_{\rm
Edd,c}=-0.5_{-0.3}^{+0.4}$, $\log f_{\rm Edd,s}=-0.4_{-0.3}^{+0.3}$,
$\log (\abbh/\rg) =2.5_{-0.2}^{+0.2}$ (here $\rg=G\mbh/c^2$), $\log
f_{\rm r,s}=-0.5_{-0.4}^{+0.4}$, $E_{\rm B-V}=0.10_{-0.03}^{+0.03}$,
and the orbital period $\simeq 1.2^{+0.2}_{-0.1}$\,yr. Note these
values for the model parameters are somewhat different from those from
the best fit with minimum $\chi^2$-value.

The X-ray emission of AGNs and quasars is normally emitted from a
corona structure above the inner disk area in the standard
disk--corona model \citep{HM1993,Dai10}.  Here in Mrk\,231, the outer
circumbinary disk is truncated at a radius substantially larger than
several hundreds of the Schwarzschild radius of the primary SMBH, for
which the disk is too cold and a corona probably cannot be established
above it to emit X-rays significantly. Therefore, the X-ray emission
can only arise from the mini-disk or its associated corona around the
secondary SMBH in the BBH--disk accretion model.  Since the secondary
SMBH is smaller than the primary one by a factor of $\sim 1/q \sim
38$, it is almost guaranteed that the X-ray emission from this system
is much weaker than that from a single SMBH--disk accretion system
with an SMBH mass of $\mbhp+\mbhs$ and an accretion rate of $f_{\rm
Edd,c} \dot{\mbh}_{\rm Edd,c}$ (or $f_{\rm Edd,s} \dot{\mbh}_{\rm
Edd,c}$). The best fit suggests that the observed X-ray luminosity of
Mrk\,231 in the $2$-$10$\,keV band is about $1\%$ of the
bolometric luminosity from the mini-disk, which is well consistent
with those for normal AGNs and quasars \citep{Hopkins07}.  This solves
the mystery of the intrinsic X-ray weakness of Mrk\,231.

\section{Model Implication and Discussions}
\label{sec:discussions}

According to the model fits, the intrinsic continuum and SED of Mrk\,231 is
significantly different from the canonical ones of normal quasars. The deficit
of intrinsic UV emission will lead to substantially weaker broad line emissions
compared with those of normal quasars. However, a number of broad emission
lines, such as H$_\alpha$, H$_\beta$, are evident in the optical spectrum of
Mrk\,231. The ratio of the H$\alpha$ flux to the H$\beta$ flux is roughly $3$,
consistent with those of normal quasars.  Here we check whether the total
number of ionizing photons emitted by the central source is large enough to
balance the total number of recombinations occurred in the broad line region.
The total number of H$_\beta$ photons is directly related to the number of
ionizing photons as
$
L_{{\rm H}_\beta} = \frac{\Omega}{4\pi} \frac{\alpha^{\rm eff}_{{\rm
H}_\beta}({\rm H}^0, T)} {\alpha_{\rm B}({\rm H}^0, T)} h\nu_{{\rm H}_\beta}
\int^{\infty}_{\nu_0} \frac{L_{\nu}}{h\nu}d\nu,
$
where $\Omega/4\pi$ is the covering factor, $\alpha^{\rm eff}_{{\rm
H}_\beta}/\alpha_{\rm B}$$\sim  1/8.5$ is the number of H$_\beta$ photons
produced per hydrogen recombination, and $\nu_0=c/912$\,{\AA}, $\nu_{{\rm
H}_\beta}=c/4861$\AA\ \citep{OF06}. Subtracting the best-fit continuum from the
observed spectrum from 4700\AA\, to 5800\AA\,, we obtain H$_{\beta}$ luminosity
by integrating the residual spectrum.  Using the above equation, we find
$\Omega/4\pi \sim 0.29$, which is fully consistent with the typical range of
$\Omega$ for AGNs/quasars and suggests that the ionizing photon emission from
the mini-disk around the secondary SMBH is sufficient to produce the optical
broad emission lines, such as H$\alpha$ and H$\beta$. 

The inclination angle of the disk in Mrk\,231 could be larger, e.g.,
$\cos i \sim 0.5$, since Mrk\,231 is an FeLoBAL (though not a normal
FeLoBAL in the BBH--disk accretion scenario) and FeLoBALs were
suggested to have a larger $i$ and a smaller $\cos i$ \citep{GM95}. By
alternatively setting $\cos i=0.5$, the best fit obtained from the
MCMC fitting suggests $\log(\mbh/\msun)=8.5_{-0.2}^{+0.2}$, $\log q =
-1.9_{-0.2}^{+0.2}$, $\log f_{\rm Edd,c}= -0.5_{-0.4}^{+0.3}$, $\log
f_{\rm Edd,s}=-0.3_{-0.2}^{+0.2}$, $\log (\abbh/\rg)
=2.5_{-0.2}^{+0.2}$, $\log f_{\rm r,s}= -0.5_{-0.3}^{+0.3}$, and
$E_{\rm B-V}=0.07_{-0.02}^{+0.02}$.  With this BBH configuration, the
orbital period is $1.6^{+0.3}_{-0.2}$ year and $\Omega/ 4\pi \simeq
0.5$. It appears that the results are qualitatively consistent with
those obtained by assuming $\cos i =0.8$. By relaxing the inclination
angle as a free parameter, we find that our results are not affected
significantly.

\begin{figure}[!ht]
\centering
\includegraphics[scale=0.3,angle=-90]{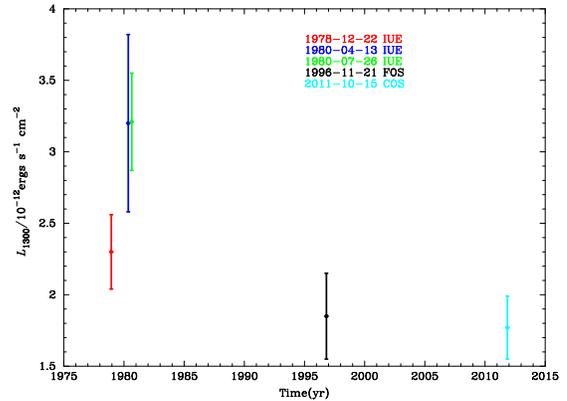}
\caption{
UV continuum flux at $1300$\AA\ ($L_{1300}$) for Mrk\,231 at different
epochs.  Different points are obtained from the UV spectra obtained by
IUE and HST at different epochs as marked in the figure, smoothed over
a wavelength range of $20$\AA\ around $1300$\AA\ at the rest frame.
The bars associated with each point represent the 1-$\sigma$ standard
deviation of the measurement.  }
\label{fig:f6}
\end{figure}

Numerical simulations suggest that the continuum emission from a
BBH--disk accretion system may vary periodically due to the change of
the material infalling rate from the circumbinary disk to the inner
mini-disk(s) \citep[e.g.,][]{Hayasaki08}, though the variation for
those systems with small mass ratios (less than a few percent) may not
be significant \citep[e.g.,][]{Farris14b}. Several UV observations of
Mrk\,231 have been obtained by {\it Hubble Space Telescope} (HST) over the years. The UV continuum of
the HST COS observations by \citet{Veilleux13} is roughly consistent
with the HST G160H observation by \citet{Gallagher02}, while it is
perhaps a factor of two higher than the HST G190/270H observation by
\citet{Smith95}. As seen from Figure~\ref{fig:f6}, earlier
observations by IUE \citep{HN87} show that the continuum ($\lambda
L_\lambda$) levels of Mrk\,231 at $1300$\AA\ $\sim 2.30, 3.20,$ and $
3.21\times 10^{-12}{\rm erg\ cm^{-2}\ s^{-1}}$ at three different
epochs, and later on they are about $1.85$ and $1.77 \times 10^{-12}
{\rm erg\ cm^{-2}\ s^{-1}}$ measured by the HST COS \citep{Veilleux13}
and FOS \citep{Gallagher02}.  The probability for those observations
to be consistent with a constant (or no variation) is $0.00013$
according to the $\chi^2$-statistics, which suggests that the UV
emission over the past several decades does vary, confirming the claim
by \citet{Veilleux13}.  However, these observations are not sufficient
for investigating the expected (quasi-)periodical variations as done
for the recently reported BBH candidate PG\ 1302-102 \citep{Graham14}
because (1) the total number of the observations is small; and (2) the
intervals between some of these observations are too large compared
with the expected orbital period.  Although there are more optical
observations of Mrk 231, the number of observations at each individual
band is limited and not sufficient for an analysis on the
(quasi-)periodical variation.

Note that the periodical variation of the BBH candidate PG 1302-102
\citep{Graham14} has an amplitude of only $\sim 0.14$ magnitude (or
$\sim 14$\%), which suggests that the amplitude of the expected
periodical variations of some BBH-disk accretion systems may be small.
Future intensive monitoring the UV continuum emission of Mrk\,231 may
reveal the periodical variation, which would confirm the BBH
hypothesis and be useful to further constrain the dynamical interplay
between the BBH and its surrounding accretion flow.

It has been shown that the polarization fraction of the optical-to-UV
continuum of Mrk\,231 depends on frequency and the peak of the
polarization is around the wavelengths $\sim 3000$\AA\,
\citep{Smith95}. This dependence is probably due to the scattering
clouds distributed asymmetrically about the illuminating source
\citep{Smith95}. In the BBH scenario, the blue photons around
$3000$\AA\ are mainly emitted from the inner edge of the circumbinary
disk, where the disk may be puffed up because of the accumulation of
accreting hot material there, which may lead to more significant
scattering of those blue photons emitted from that region and thus a
high polarization fraction; while those photons at the FUV and optical
bands may experience less scattering as they are away from the inner
edge of the circumbinary disk. This may explain the frequency
dependence of the polarization fraction, qualitatively;
\citet{Veilleux13} and \citet{Leighly2014}, likewise, also provided
qualitative explanations on these polarization measurements in their
competing models.

The standard accretion disk model adopted in the fitting is simple,
and the torque due to the BBH on the outer disk is not considered.  As
shown in the Appendix, we adopt the model by \citet{Rafikov13} to fit
the continuum, in which both the internal viscosity and the external
torque by the BBH on the circumbinary disk are considered, and we find
that there are no significant differences in the constraints on the
model parameters.

According to the best fit, the infalling rate from the circumbinary
disk to the inner disk(s) is smaller than the accretion rate of the
circumbinary disk by a factor of $\sim 30$. Currently it is not clear
whether such a small infalling rate can be realized in BBH--disk
accretion systems. For close BBH--disk accretion systems, a number of
simulations and analyses suggested that the infalling rate into the
gap or the central cavity is substantially smaller than the accretion
rate at the outer boundary of the circumbinary disk because of the
tidal barrier by the central BBH (e.g., the 1D simulations by
\citealt{MP05}, the 3D simulations by \citealt{Hayasaki07}, or the
simple arguments by \citealt{Rafikov13}). However, some recent
simulations showed that the infalling rate may not be significantly
suppressed by the tidal barrier (e.g., \citealt{Noble12};
\citealt{MM08}) and it may be even comparable to the accretion rate in
the case with a single central MBH (\citealt{Farris14b};
\citealt{Shi15}). Considering that all those simulations are quite
idealized, it is also not clear whether the accretion onto the central
BBH is really significantly suppressed or not.  Future more realistic
simulations may help to demonstrate whether the small infalling rate
found for the BBH-disk accretion system in Mrk\,231 is possible.

The spectral features of Mrk\,231 were recently proposed to be
explained by either (1) a hybrid model with both a young star burst
(for the UV continuum) and an obscured quasar (for the optical
continuum) \citep[see][]{Leighly2014} or (2) an absorption model with
extremely large dust reddening at the NUV band but small reddening at
the FUV band \citep[see][]{Veilleux13}.  In the first model, the FUV
emission is dominated by the young star burst with an age of $\sim
100$\,Myr and therefore would have no significant short time-scale
variations, which appears inconsistent with the observations.
\citet{Veilleux13} also argued that the broad asymmetric Ly$\alpha$
line cannot be produced if the FUV emission is from an extended star
burst. While in the second model, a small fraction of FUV emission is
leaked through the obscuring material, and the UV variability will
depend on the stability of the leaking holes.  Future monitoring
observations of Mrk\,231 at the FUV and NUV bands, and simultaneous
multi-wavelength observations will be helpful to confirm the BBH--disk
accretion explanation and further constrain the orbital evolution of
the BBH in the core of Mrk\,231.

\section{Conclusions}
\label{sec:condis}

In this paper, we show that various unique features in the
optical-to-UV spectrum and the intrinsic X-ray weakness of Mrk\,231
can all be well explained, if a pair of SMBHs exists in the core of
Mrk\,231, with the masses of the primary and the secondary SMBHs as
$\sim 1.5\times 10^{8}\msun$ and $4.5\times 10^6\msun$, respectively.
The existence of a BBH in Mrk\,231 is compatible with its disturbed
morphology and tidal features, which indicates a merger event in the
past. (Note that the secondary SMBH is rather low in mass; however, it
should be able to sink down to the center because the stars initially
associated with it enhance the dynamical friction
\citealt[see][]{Yu02}.) The semimajor axis of this BBH is $\sim
590$\,AU, about $190$ times of the Schwarzschild radius of the primary
SMBH, and its orbital period is just $\sim 1.2$ year, relatively
short among the few known BBH candidates \citep{Valtonen08, Graham14},
which makes it an ideal system to study the dynamics of BBH systems.
Such a BBH emits gravitational wave on tens of nanohertz, and the
change rate of its orbital period due to gravitational wave radiation
is about $40$\,seconds per orbit. This BBH might be a target for
gravitational wave studies in future.

The orbit of such a BBH system decays on a timescale of a few times of
$10^5$ year due to gravitational wave radiation and the torque of the
circumbinary disk \citep{Haiman09}, which is not too small compared
with the lifetime of quasars (a few times $10^7$ to $10^8$ year; see
\citealt{YT02,Shankar13,YL08}).  The majority of quasars are believed
to be triggered by mergers of galaxies and consequently involve
mergers of SMBHs \citep{Volonteri03}, and those BBH systems with mass
ratio in the range of a few percent to 1 may lead to a notch in the
optical-to-UV continuum emission if their semimajor axes are in the
range of a few hundreds to about one thousands gravitational radii
\citep{Roedig14}, which correspond to orbital decay timescales of
$10^5$--$10^6$ year \citep{Haiman09}.  Therefore, the occurrence rate
of active BBH systems, with deficits in the optical-to-UV emission,
may be roughly a few thousandths to about one percent among quasars
\citep{Yan14}. Our analysis of Mrk\,231 demonstrates the feasibility
of finding BBH systems by searching for the deficits in the
optical-to-UV emission among the spectra of quasars, a new method
proposed by a number of authors \citep{Farris14, GM12,Sesana12,
Yan14}, in addition to those current practices by searching for the
kinematic and image signatures of BBHs among AGNs/quasars \citep{BL09,
Valtonen08, Graham14, Liu14, Gaskell96, Bogdanovic08, Tsalmantza11,
Eracleous12, Ju13, Rodriguez06, Popovic12}.

%\noindent {\bf Acknowledgments}
\acknowledgements
We thank Zheng Zheng for finding a typo in an earlier version of the paper.
This work was supported in part by the National Natural Science
Foundation of China under grant nos. 11273004 (Q.Y.), 11103029 (C.Y.),
\,11373031 and 11390372 (Y.L.), and the Strategic Priority Research
Program ``The Emergence of Cosmological Structures'' of the Chinese
Academy of Sciences, Grant No. XDB09000000 (Y.L.). X.D. is supported
by the NSF grant AST-1413056.

\appendix
\label{Appendix}

The structure of the circumbinary disk may be different from that of a
standard thin disk, especially at the region close to the inner edge,
since the torque raised by the central binary may lead to gas
accumulation there.  According to \citet{Rafikov13}, the evolution of
the circumbinary disk under the actions of both the internal viscosity
and the external torque can be described by a simple equation
\be
\frac{\partial \Sigma}{\partial t}=-\frac{1}{r}
\frac{\partial}{\partial r}\left[ \left(\frac{\partial l} {\partial
r}\right)^{-1}\frac{\partial}{\partial r}
\left(r^3\nu\Sigma\frac{\partial\Omega} {\partial
r}\right)+\frac{2\Sigma\Lambda}{\Omega}\right].
\label{eq:evSigma}
\ee
Here $t$ is the evolution time, $\Lambda$ is the external torque per
unit mass of the disk due to the central BBH, $\Sigma$,  $\nu$, and
$r$ are the surface density, kinematic viscosity, and radius of the
disk, respectively, and $l=\Omega(r)r^2$ is the specific angular
momentum of gas material on a circular orbit with a radius $r$.  At
each radius, the accretion disk properties can be characterized by the
viscous angular momentum flux
\be
\FJ\equiv -2\pi\nu\Sigma r^3\frac{d\Omega}{dr}=3\pi\nu\Sigma \Omega
r^2,
\label{eq:Fnu}
\ee where the last equality is valid for a near Keplerian disk with
$\Omega \simeq (G\mbh/r^3)^{1/2}$ and $\mbh=\mbhp+\mbhs$.

\begin{figure}
\centering
\includegraphics[scale=0.5,angle=-90]{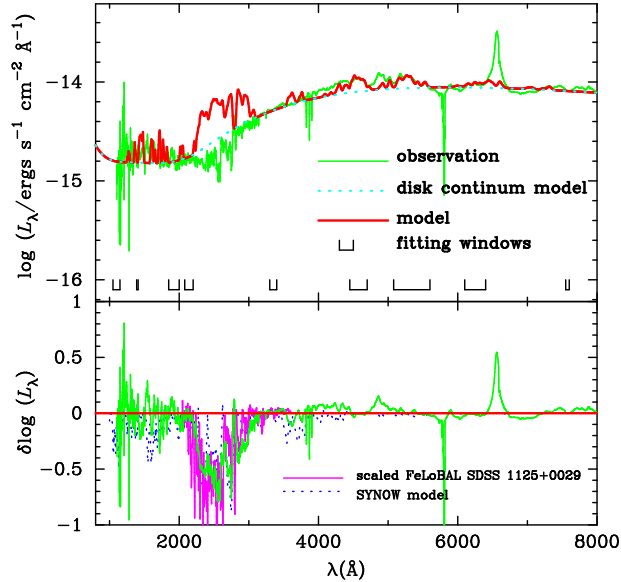}
\caption{
The optical-to-UV spectrum of Mrk231 and the model spectrum.  Legend
is similar to that for Fig.~\ref{fig:f3}, except that the torque at
the inner edge of the circumbinary disk is considered in the model
fitting to the continuum emission.
}
\label{fig:f7}
\end{figure}

Since $\Lambda$ in equation (\ref{eq:evSigma}) is significant only in
a narrow annulus at the inner edge of the disk, we can assume that
$\Lambda=0$ outside of some radius $r_{\Lambda}$, which is not too
different from $r_{\rm in}$.  Outside $r_{\Lambda}$, the disk evolves
according to Equation~(\ref{eq:evSigma}) with $\Lambda=0$. The
solution of this equation can be obtained by the standard approach
introduced in Lynden-Bell \& Pringle (1974) \cite{LBP74}, if the inner
boundary condition (IBC) is assumed to be torque free. However, the
central BBH exerts a torque on the inner boundary of the circumbinary
disk, which can be approximated by $\FJ(r_{\rm in})\approx -dL_{\rm
BBH}/dt=-L_{\rm BBH}v_{\rm BBH}/2\abbh$, where $L_{\rm BBH}=q
\mbh(G\mbh\abbh)^{1/2}/(1+q)^2$ is the orbital angular momentum of the
BBH, and $v_{\rm BBH}\equiv d\abbh/dt$ is the inspiralling speed of
the BBH.  According to Rafikov (2013) \cite{Rafikov13}, this IBC may
be written as
\bea
\frac{\partial\FJ}{\partial l}\Big|_{r=r_{\rm in}}=\dot M(r_{\rm in})=
\chi\dot M_\infty,
\label{eq:BC}
\eea
where $\chi$ is assumed to be a constant ($\le 1$).  By this setting,
$\dot M(r_{\rm in})$ can be substantially smaller than $\dot
M_\infty$, and disk material may gradually accumulate near the inner
boundary if $\chi <1$.  We set  the accretion rate at the outer
boundary as $\dot M_\infty$. A self-similar solution can be obtained
for the evolution of the circumbinary disk once the initial separation
of the BBH is given [i.e., $\abbh(t=0)=a_{\rm BBH, 0}$]. With the
solution of $\FJ(r, t)$, the effective temperature of the disk can be
obtained as
\bea
T_{\rm eff}(r,t)=\left[ \frac{3}{8\pi}\frac{\FJ(r,
t)\Omega}{\sigma_{\rm B} r^2}\right]^{1/4}.
\label{eq:T_gen}
\eea
By integrating the multi-color black body emission over the whole
circumbinary disk, we obtain the continuum flux as
\be
F_{\lambda, \rm c}=\int^{r_{\rm out,c}}_{r_{\rm in,c}}
\frac{2\pi r}{D^2_{\rm l}}
\frac{2 hc^2
\cos i /\lambda^5}{\exp[{\hp c/\lambda k_{\rm B}\teff(r)}]-1}dr.
\label{eq:F1A}
\ee
Here $D_{\rm l}$ is the luminosity distance of Mrk\,231, $\hp$ is the Planck constant, $k_{\rm B}$ is the Boltzmann
constant, ${r_{\rm in,c}=\abbh(t)/(1+q)+R_{\rm H}}$, $r_{\rm out,c}$
is set to be $10^5GM_{\bullet}/c^2$, and $\abbh(t)$ is the semimajor
axis of the BBH at time $t$. The evolution of $\abbh(t)$ is controlled
by
\be
\frac{d \abbh}{dt}=-\frac{\abbh}{t_{\rm GW}}-\frac{\abbh}{t_{\rm J}},
\ee
where $t_{\rm GW}=[5(1+q)^2/64q](\rg/c)(\abbh/\rg)^4$ is the
gravitation wave radiation timescale \cite{Peters64}, $\rg =
G\mbh/c^2$ is the gravitational radius, and $t_{\rm J} \equiv L_{\rm
BBH}/2\FJ$ is the characteristic timescale for the BBH orbit shrinking
due to the coupling to the circumbinary disk. If we adopt this model
to describe the circumbinary disk and adopt the standard thin disk
model to describe the mini-disk, we can also fit the continuum of
Mrk\,231 by the MCMC technique. Under this approach, $f_{\rm Edd,s}
=\chi f_{\rm Edd,c}/q$. The best fit is shown in Figure \ref{fig:f7},
and the constraint on $a_{\rm BBH,0}$ and $t$ are $355\rg$ and
$2.1\times 10^4$ year, respectively. Other best-fit parameters are
$\mbh=10^{8.2}\msun$, $f_{\rm Edd,c}=0.4$, $q=0.02$, $\chi=0.04$
(corresponding to $f_{\rm Edd,s}=0.7$), $\abbh=343\rg \sim 540$\,AU
(corresponding to $r_{\rm in,c}=402 \rg$, $f_{\rm r,s}=0.12$), and
$E_{\rm B-V}=0.14$.  These results are roughly consistent with those
shown in Figure~\ref{fig:f3}, which are obtained without considering
the torque of the BBH on the inner boundary of the circumbinary disk.

The torque due to the outer circumbinary disk on the mini-disk around
the secondary SMBH and the shock induced by the infall streams onto
the mini-disk \citep{Lodato09, Kocsis12, Farris14, Roedig14}, which
are omitted in this study, may also introduce some errors to the
fitting.  Although many simulations and analysis suggest a flux
deficit in the continuum emission from a BBH-disk accretion system
because of the gap (or hole) in the disk, including the one by
\citet{Roedig14} with consideration of the shock heating hot-spot on
the inner mini-disk(s), one new simulation by \citet{Farris14}
suggested that there may be no strong flux deficit in the continuum
emission of a BBH-disk accretion system by considering the emission
from the infalling streams.  However, only a specific case with a mass
ratio of 1 is considered in \citet{Farris14} and the stream emission
therein is approximated as a thermal emission without considering
inverse Compton scattering and radiative transfer. Therefore, their
results might not be applicable to other cases and need to be improved
(see the discussion in \citealt{Farris14}).  \citet{Roedig14} argued
that the shock induced by the infalling streams on the mini-disk(s)
may enhance the X-ray emission at $\gtsim 100$\,keV, which does not
affect the X-ray emission with energy significantly less than
$100$\,keV and thus does not affect our estimate on the $2-10$\,keV
emission.

\end{document}